# Anisotropic magnetoresistance of bulk carbon nanotube sheets

E. Cimpoiasu, G. A. Levin, B. White, and D. Lashmore

Department of Physics, United States Naval Academy, Annapolis, MD, 21412 USA
Air Force Research Laboratory, Wright-Patterson AFB, OH, 45433 USA
Nanocomp Technologies, Inc., Concord, NH 03301

Abstract: We have measured the magnetoresistance of stretched sheets of carbon nanotubes in temperatures ranging from 2 K to 300 K and in magnetic fields up to 9 T, oriented either perpendicular or parallel to the plane of the sheets. The samples have been partially aligned by post-fabrication stretching, such that the direction of stretching was either parallel or perpendicular to the direction of applied electric current. We have observed large differences between the magnetoresistance measured under the two field orientations, most pronounced at the lowest temperatures, highest fields, and for the laterally-aligned sample. Treatment of the sheets with nitric acid affects this anisotropy. We analyzed the results within the theoretical framework of weak and strong localization and concluded that the anisotropy bears the mark of a more unusual phenomenon, possibly magnetically-induced mechanical strain.

**PACS**: 73.43.Qt, 73.63.Fg, 75.80.+q

# 1. INTRODUCTION

Carbon nanotubes (CN) hold unique properties, like very low density, high thermal and electric conductivity, and high mechanical strength.[1] When fabricated in sheets, these unique properties make them very attractive for many applications, including consumer electronics, sensing, or conducting sheets for shielding. The charge transport mechanisms are the key factors in the material's performance. Understanding the parameters that influence the charge transport and mapping the boundaries of applicability are therefore crucial. Many studies have been concerned with the electrical properties of carbon nanotubes networks. In particular, magnetoresistance (MR) has been used as a sensitive tool to assess the transport phenomena in several CN networks, doped or not, fabricated in films, or sheets, of uncontrolled mixtures of metallic and semiconducting CNs.[2-7] However, only few reports touch on the anisotropic features of magnetoresistance.[2, 7-9] This is a topic highly relevant to CN systems, as the nanotubes are inherently anisotropic, but is especially relevant to anisotropic networks, such as partially-aligned networks or ropes.

Here we report a detailed study of the anisotropic properties of the magnetoresistance of CN sheets that have been stretched after fabricaton, such that the direction of stretching was either parallel or perpendicular to the direction of applied electric current. The magnetic field has been applied either parallel of perpendicular to the plane of the sheet. The magnetoresistance can be resolved in two terms, one negative and the other positive. Both MR terms exhibit anisotropy with respect to the field orientation, the direction of CN alignment, and doping with nitric acid. Many of the previous reports on similar systems explained the MR results in terms of variable-range hopping for the nonmetallic networks and weak localization for the metallic ones.[2-7] We have performed a similar analysis, but focused mainly on the anisotropic properties. We concluded that in the case of our samples the anisotropy is a result of a different phenomenon, generated by the presence of the iron-based catalyst nanoparticles. We suggest that the nanoparticles are magnetized in the presence of the magnetic field and produce a $B$-dependent strain, which in turn affects the value of resistance.

## 2. EXPERIMENTAL

The samples were prepared from bulk CN sheets produced by Nanocomp Technologies using a chemical vapor deposition (CVD) process previously described.[10,11] Briefly, an alcohol-based fuel containing ferrocene as the catalyst source is introduced into a tube furnace using hydrogen or helium as a carrier gas. The fuel components decompose and flow through the furnace exiting out of the opposite end as a casing (or sock) composed of CNs. The CN sheet is collected on a rotating, translating belt. The output of this process is a final sheet comprised of hundreds of layers of CN, with a total thickness of ~ 60 μm.

Two categories of samples have been prepared. The first category, denoted "undoped" in that it did not suffer any additional chemical treatment, was prepared by soaking a 2 cm by 8 cm strip in ethanol and then slowly stretching it 22% between two clamps, after which it was allowed to dry under tension. Two test samples (2.5 mm by 10 mm) were cut from this strip, such that the long axis was parallel to the stretch direction (sample UD-long) and perpendicular to the stretch direction (sample UD-lat).

The second category was termed "doped" in that a 2 cm by 8 cm strip was subjected to nitric acid treatment. The treatment was performed by soaking the sample in 8 M nitric acid for 30 minutes, after which it was rinsed with water until neutral, and air dried under tension. A final drying step was carried out at 100 °C for one hour to assure the complete removal of water. The sample was re-wet with water and slowly stretched by 18% between two clamps and allowed to dry under tension. Two test samples (2.5 mm by 10 mm) were cut from the strip such that the long axis was parallel to the stretch direction (sample D-long) and perpendicular to the stretch direction (sample D-lat).

Scanning electron images of the D-long sample were acquired with an Amray 3300FE microscope and are shown in Fig. 1. The stretching direction is also indicated using an optical image of the actual sample. The CN network is a mixture of individual nanotubes and bundles of tubes of various diameters, decorated with catalyst nanoparticles. The stretching does not produce an obvious alignment of the bundles and strands, as they appear to mingle in all directions. However, a closer inspection [see Fig. 1(b)], especially of the direction of the largest diameter strands, suggests an overall positioning consistent with the direction of stretching.

The electrical resistance of the four CN sheets was measured using the four-point method. The geometry of the setup is shown in the cartoon of Figure 2. The stretch directions are indicated with white arrows. The pads, ~ 1 mm in width, were electrodeposited by first depositing a layer of nickel followed by a layer of copper. Thin gold wires were then silver-bonded to the pads and used to connect the sample to either a resistance bridge or a SR830 lock-in preamp. An electric current of 100 μA was applied along the long axis of the samples as shown in Fig. 2. We have used a Quantum Design PPMS to change the temperature of the sample from room temperature $T = 300$ K down to 2 K and the magnetic field, *B*, up to 9 T. The resistance-magnetic field curves were recorded for two different field orientations: one perpendicular to the plane of the fabric, $B\perp$, and one parallel to the plane of the fabric, $B_{\parallel}$ (see drawing in Fig. 2). It can be seen that at all times the electrical current *I* is perpendicular to the magnetic field.

## 3. ZERO-FIELD RESISTANCE

Figure 2 shows the temperature dependence of the square resistance *R* . The effect of doping and alignment on the *R* -*T* curves is very pronounced. First, doping reduces the square resistance and modifies the temperature behavior, consistent with the improvement of the nanotubes intrinsic conductivity and the intertube junctions.[12] Specifically, the undoped samples show nonmetallic *T*-dependence over the entire temperature range, 2K-300 K, with a hint of saturation above room temperature. The doped samples exhibit a much weaker nonmetallic *T*-dependence up to ~ 250 K where it turns metallic. The resistivity ratio α = *R* (2K)/ *R* (40K) is a quick indicator of transport state of the samples.[2] The α-values are 1.45, 1.48 for the doped samples and 3.4, and 3.23 for the undoped ones (longitudinally and laterally-stretched, respectively). These values are consistent with improved conductivity in the doped samples, and correspond to the critical regime according to the classification of Ref. [2], where values of α > 4 corresponds to the insulating regime, α < 1.5-2 to the metallic regime, and 2 < α < 4 to the critical regime of the metal-insulator transition.

The effect of alignment is quite interesting. For both doped and undoped samples, *R* is anisotropic in that the square resistance is lower when the current flows in the direction of the stretch. This result seems to imply that the irreversible stretching of the fabrics shortens the average path of the carriers through the CN network by straitening the nanotubes along the direction of the stretch and reducing the number of intertube junctions.   Moreover, the temperature dependences are strikingly similar for the two kinds of alignments as seen from the log-log plot of Fig. 2, where the two curves appear to be shifted by a constant.  The $R$ -T dependences are, however, not perfectly the same.  The temperature dependences of $R_{lat}/R_{long}$ for both the doped and undoped samples are illustrated in the inset of Fig. 2.

The details of the temperature dependence of the zero-field transport are shown in Fig. 3. For 26 K< T <196 K, the $T$-dependence of the conductance of the undoped samples can be described as [Fig. 3(a)]:

$$G_{lat,long}(T) \sim T^{1/2}$$
(1)

Below 26 K the insulating behavior is stronger and the samples transition to the strong localization regime. This becomes more evident when we consider the reduced activation energy defined as

$$W(T) = -\frac{d \ln R(T)}{d \ln T} = \frac{d \ln G}{d \ln T}.$$

In the case of variable range hopping, $W(T) \sim s(\frac{T_0}{T})^s > 0$, where the exponent $s$ = ½, ⅓, or ¼ depending on the dimensionality of the substance and the type of interaction.[2,13] The inset to Fig. 3(a) shows $W(T)$ for the undoped laterally-stretched sample.  The fractional $T$-dependence $W \sim T^{-0.27}$ is consistent with variable range hopping.[2-5]

The $T^{1/2}$ dependence of the conductance over a relatively large temperature range was previously observed as well and was interpreted in terms of weak localization and electron-electron interaction effects.[2,6]   Indeed, the characteristic corrections to conductivity obtained for 3D systems within the weak localization theory which takes into account electron-electron interactions are written as: [14]

$$\Delta\sigma(T) = \frac{e^2}{\hbar} \frac{1}{\pi^3} \frac{T^{p/2}}{a} + \frac{e^2}{4\pi^2 \hbar} \frac{1.3}{\sqrt{2}} (\frac{4}{3} - \frac{3}{2} \tilde{F}_\sigma) \frac{\sqrt{T}}{D} \qquad (2)$$

where $p/2$ is the temperature index of the phase coherence length $L_\Phi \sim aT^{-p/2}$ and is dependent on the dominant collision mechanism, $D$ is the diffusion coefficient, and $\tilde{F}_\sigma$ is the interaction parameter. However, the network of carbon nanotubes is a very different system than the bulk substance for which the corrections given by Eq. (2) were derived. Therefore, the formal similarity between Eqs. (1) and (2) should be interpreted with caution.

Figure 3(b) shows the conductance data for the doped samples. Below 60 K, the conductances curves have a logarithmic dependence:

$$G_{lat,long}(T) \propto \ln(T/T_0) \qquad (3)$$

We have confirmed this behavior by plotting the derivative $dG/dlnT$ vs. $T$ in the inset to Fig. 3(b). The curve is $T$-independent up to ~60 K. This kind of dependence has been observed before and has been attributed to 2D transport.[3, 5] Indeed, the logarithmic dependence of conductivity appears as a correction in the 2D theory of weak localization theory which takes into account electron-electron interactions:[5, 14]

$$\Delta\sigma(T) = \frac{e^2}{2\hbar\pi^2}[\alpha p + (1 - \frac{3}{4}\tilde{F}_\sigma)]\ln(\frac{T}{T_0}) \qquad (4)$$

However, it is rather obvious that the network of nanotubes can hardly be described as a 2D system. Alternatively, Ref. [2] points out that logarithmic $T$-dependence of resistivity is characteristic to materials with α ~ 1.5, which separates the metallic (sublogarithmic dependence) from the nonmetallic (superlogarithmic dependence) materials. This observation is especially relevant to our doped samples, for which α=1.45 and 1.48.

It is probably more instructive to pay attention to the value of the characteristic temperature $T_0$ in Eq. (3). Extrapolating the data in Fig. 3(b), we obtain $T_0$ of approximately 30 mK. The logarithmic dependence must break down somewhere in the vicinity of this temperature. One can think of several scenarios for such breakdown. For example, one can suggest that the conductivity will turn more insulating in the vicinity of $T_0$. However, the extremely small value of the energy scale associated with $T_0$ makes this scenario unlikely. For instance, in the undoped samples the transition to the variable range hopping takes place at much higher temperature, $T = 26$ K. There are only few plausible types of transitions that can be associated with such extremely small energy scale. A superconducting transition is one of them. One can speculate that the characteristic logarithmic dependence of conductivity may indicate a possibility of superconducting transition at subkelvin temperatures. This hypothesis can be confirmed experimentally if similar doped samples can be cooled down to the level of $T_0$.

## 4. MAGNETORESISTANCE

A more sensitive tool to investigate the transport properties is the magnetoresistance (MR). Figure 4 shows the percentage magnetoresistance $\Delta R/R=[R(B)-R(0)]/R(0)\cdot100\%$ for the laterally-stretched, undoped [Fig. 4(a)] and doped [Fig. 4(b)] samples, in field applied perpendicular to the plane of the fabric. The data for the longitudinally-aligned samples or other orientations is similar in the overall features to the data shown in Fig. 4.

The MR is clearly nontrivial, with several noteworthy features consistent with features previously reported on similar networks:[2,4,5,7]

(a) A negative MR component $\Delta R/R_{neg}$ is superimposed on a positive term $\Delta R/R_{pos}$. The total magnetoresistance is thus equal to the sum of the two terms $\Delta R/R = \Delta R/R_{neg} + \Delta R/R_{pos}$.

(b) The field dependences of the two terms are very different. The negative term tends to saturate at high fields for all temperatures. The positive term is generally quadratic $\Delta R/R_{pos} \sim \gamma B^2$, with the exception of the lowest temperatures. Interestingly enough, although the positive term is very small for the doped samples, it also shows signs of deviation from the quadratic dependence at T = 2 K.

(c) Chemical treatment with nitric acid greatly diminishes the positive term.

One other feature of our results, which is also the most puzzling, is the anisotropy of the MR. The total MR and each of the components are anisotropic with respect to the direction of the applied field and the direction of current relative to the fiber alignment. In order to understand the source of this anisotropy, further analysis can be accomplished by separating the two contributions to MR. As an example, Fig. 5(a) shows MR of the laterally-stretched undoped sample at $T = 5$ K for two field orientations, parallel and perpendicular to the plane. We first fitted the $\Delta R/R$ curves in the higher field region with a $\sim B^2$ dependence. Then, the results of the fit shown by the continuous lines in the figure and labeled $\gamma_{\parallel} B^2$ and $\gamma_{\perp} B^2$ were subtracted from the data. The remaining negative contributions to MR, labeled $\Delta R_{\parallel,neg}$, and $\Delta R_{\perp,neg}$ are also shown. For comparison, the data for the laterally-stretched doped sample are shown in Fig. 5(b).

A. The negative of component of MR. Quantum interference effects.

The negative components of MR of both the doped and the undoped samples are shown in the inset to Fig. 5(b) for the two orientations of the applied field. The overall field dependence is very similar for all four curves and the values of the field at which these contributions saturate are within 15% of each other. Considering how different the temperature dependence of the resistivity for doped and undoped samples is (see Figs. 2 and 3), this similarity of the MRs indicates a common charge transport.

Let us further explore the specific $H$-dependence of this negative term. Figure 6 shows the details of the field dependence for the undoped laterally-stretched sample at $T = 10$ K, under two different field orientations. This set is representative for the analysis we performed for all the other samples and temperatures. The negative component of MR is plotted vs. $B^{1/2}$, although in this range of fields one cannot reliably distinguish between the $B^{1/2}$ and the $ln(B)$ dependences.

There are two important parameters one can extract from this data. The transition from the initial $\sim B^2$ dependence at low field to the $B^{1/2}$ or $ln(B)$ dependences taking place at a certain temperature dependent field $B_\Phi$. This characteristic field can be determined from the data either as an intercept of the straight line as shown in the Fig. 6, or by fitting the low field region with a $\sim B^2$ dependence, shown in Fig. 6 by the solid curves, and defining $B_\Phi$ as a point where the fit deviates from the data. Either way we obtain very similar values of $B_\Phi$. The second important parameter is the field value at which the negative component of MR saturates. We define $B_{sat}$ as the value where the field dependence becomes practically flat, as shown in Fig. 6.

The negative magnetoresistance of the type shown in Figs. 4-6 is known to manifest itself in bulk substances as well as in the CN networks. In the bulk materials it is typically associated with quantum interference effects in self-intersecting trajectories. The 3D weak localization theory predicts that the magnetoconductivity depends on the applied magnetic field according to: [14]

$$\Delta\sigma_{WL}(B.T) = \frac{1}{12\pi^2}\frac{e^2}{\hbar}(\frac{e}{\hbar})^2 L_\Phi^3 B^2 \text{, at low fields, when } \hbar/(4eBL_\Phi^2) >> 1 \text{ ,} \tag{5}$$

and to:

$$\Delta\sigma_{WL}(B.T) = \frac{e^2}{2\pi^2\hbar}\sqrt{eB/\hbar} \times 0.605 \text{, at high fields, when } \hbar/(4eBL_\Phi^2) << 1 \tag{6}$$

In the CN networks the effect of interference must have significant differences from that in the bulk substances. Since the charge carriers are confined within the quasi one-dimensional fibers, the properties of the trajectories responsible for interference are determined by the topological properties of the network itself, its composition, and the way the individual CNs form a contact with each other. For instance, a possible contribution to the negative MR may come from irregularly shaped loops formed by the randomly assembled carbon nanotubes, assuming that the CN junctions do not change the phase coherence of the carriers. The application of magnetic field perpendicular to these self-intersecting trajectories will reduce the interference. The result is a negative MR similar, but not necessarily identical, to the MR due to self-intersecting trajectories in the bulk substances.

The transition from the initial ~$B^2$ dependence at low field occurs at the characteristic field $B_\Phi$, where the magnetic flux $A_\Phi B_\Phi$ through the area $A_\Phi \sim L_\Phi^2$ of the interfering loop is of the order of one flux quantum, $\Phi_0=h/2e$. Typically $B_\Phi$ is expressed in the form:

$$B_\Phi = \frac{\phi_0}{4\pi L_\Phi^2} \tag{7}$$

Figure 7 shows the resulting phase coherence length for the undoped [Fig. 7(a)], and doped [Fig. 7(b)] samples obtained from the experimental values of $B_\phi$ using Eq. (7). There is no significant difference between the values of the coherence length in the doped and undoped samples as evident from the data shown in the inset in Fig. 5(b). The values of the phase coherence length vary in the range from ~ 40 nm at T = 2K, to ~8 nm at the highest temperatures. Despite the noisy data, one can resolve a difference between the *T*-dependence of the phase coherence length of the laterally- and longitudinally-stretched samples, as shown in Figure 7. The straight lines are power fits $aT^{-p/2}$ to the data. The resulting *p*-values are ~ 0.64 for the longitudinally-stretched samples, and ~0.4-0.5 for the laterally-stretched samples. Although typical *p*-values are larger than unity,[6, 14] the results are consistent to other reported values on similar samples.[2]

Figure 8(a) addresses the *T*-dependence of the saturation values of the negative component of MR, $|\Delta R/R_{neg, sat}|$, for the undoped samples. All curves share a common *T*-dependence, up to multiplicative constants which account for the anisotropy. The saturation value increases with decreasing *T*, as expected, but reaches a peak at approximately 10 K. At the lowest temperatures, the anisotropy of this component of MR is reduced to a minimum.

The physical meaning of the magnetic field $B_{sat}$ at which the saturation takes place can be related to the smallest size of the loops $l_{small}$ participating in the interfering processes. In the bulk materials, it corresponds to the elastic scattering length. Similar to Eq. (7) we define $l_{small}$ in terms of $B_{sat}$ as follows:

$$B_{sat} = \frac{\phi_0}{4\pi l_{small}^2} \qquad (8)$$

Figure 8(b) shows the values of $l_{small}$ as a function of temperature. It averages approximately 4.5 nm and is temperature independent in a wide temperature range, but increases sharply below 10 K. This is the result of the strong reduction of the saturation field $B_{sat}$ [see inset to Fig. 8(b)]. The temperature independent $B_{sat}$ and, correspondingly, $l_{small}$ is a natural measure of the minimum size of the interfering paths. The increase of $l_{small}$ at low temperatures is difficult to interpret. Perhaps, at these low temperatures and relatively high fields, other effects start to manifest in the CN network, like the onset of strong localization.

### B. The positive component of MR. Strong localization.

We focus next on the nature and the properties of the positive component of the MR. The temperature dependences of the quadratic coefficients $\gamma_\parallel$ and $\gamma_\perp$ are shown in Fig. 9. Figure 9(a) presents the data for undoped samples. There are two remarkable aspects worth noting: first, the quadratic coefficients are anisotropic with respect to the direction of the magnetic field and the direction of the stretch. For instance, the largest values of $\gamma$ correspond to the current flowing in the perpendicular direction to the stretch. Also, $\gamma$ is largest when the magnetic field is applied parallel to the plane; second, the overall dependence of $\gamma$ is well described by $\gamma = m_0 + m_1 T^{-\frac{3}{2}}$ at the lowest temperatures, a temperature range where the samples transition to the strong localization regime. In the VRH transport regime, both a positive and a negative MR response is expected to appear. The positive MR is due to the shrinkage of the localization length in the direction perpendicular to the applied magnetic field. The MR due to Coulomb-gap VRH in the low $B$ limit is described by:[7]

$$\ln\frac{\rho(B)}{\rho(0)} = 0.0015(\frac{\lambda}{L_H})^4(\frac{T_0}{T})^{\frac{3}{2}}, \qquad (9)$$

where $L_H = \sqrt{\hbar/eB}$ is the magnetic length, $\lambda$ is the localization length, $T_0 = \beta e^2/k_B 4\pi\varepsilon_0\kappa\lambda$, $\kappa$ is the dielectric constant, and $\beta$ has the value 2.8 in 3D. Thus,

$$\gamma = 0.0015\lambda^4(\frac{e}{\hbar})^2(\frac{T_0}{T})^{\frac{3}{2}} \sim T^{-\frac{3}{2}}. \qquad (10)$$

The slope $m_1$ can be used to find the localization length $\lambda$. Assuming a dielectric constant of $\kappa$=60,[7] we find localization lengths of 6.8 nm, 7.5 nm, 7.0 nm, and 7.2 nm for the undoped laterally-stretched in $B \perp$, laterally-stretched in $B \parallel$, longitudinally-stretched in $B \perp$, and longitudinally-stretched in $B \parallel$, respectively. Assuming an average of 7 nm, this yields $T_0$ = 111 K. These values are in agreement with reported values of localization length on similar samples.[7] Although there is some variation in the localization length values which follows the anisotropy described above, the differences are not large enough to account for the big observed anisotropy. The main contributor is the intercept $m_0$, which is positive [see inset to Fig 9(a) for enlarged picture at the highest temperatures] for both undoped samples under $B \parallel$ orientation.

Fig. 9(b) depicts the $\gamma$ coefficients for the doped samples. The values are very small, one order of magnitude smaller compared to the $\gamma$ –coefficients of the undoped samples, and drop to zero at temperatures above 7 K. The temperature dependence is different as well, following an approximately $T^{-3/4}$ law.

C. Anisotropy of the magnetoresistance.

The most intriguing feature of our magnetoresistance results is the anisotropy of the MR with respect to the field orientation. In the undoped case, the positive component of MR is substantially greater when the field is oriented parallel to the plane than when the field is perpendicular to the plane, i. e. $\gamma_{||} > \gamma\perp$. The same $|\Delta R/R_{Dneg,lat,||}| > |\Delta R/R_{D,neg,lat},\perp|$ is true for the magnitude of the negative component, although the effect is not as pronounced. This direction of anisotropy in the undoped samples is hard to understand in the context of variable range hopping, for instance, where the field effect originates from the shrinkage of the localization radius in the plane perpendicular to the field. Indeed, a three-dimensional sample should exhibit no anisotropy if the magnetic field is applied perpendicular to the direction of the current, as it is our case. We should mention that strong localization was not the only candidate theory used in the literature to explain the observed positive magnetoresistance. Similar MR results have been attributed to electron-electron interaction (EEI)[6] and 2D spin-dependent VRH.[5, 15] However, the observed anisotropy does not favor these theories either, as they predict isotropic contributions with respect to the magnetic field orientation.

Next we discuss another possible origin of the MR anisotropy with respect to field orientation, namely, magnetoelastic effects. As mentioned in Section II, the samples of CN fabric contain a substantial amount of iron-based nanoparticles. In the presence of magnetic field, the magnetic moments of the nanoparticles align along the field. The alignment results in strong magnetic forces distributed throughout the entire volume of the fabric. Indeed, we have observed experimentally that if the samples are not attached to the substrate, they will turn and reposition themselves so that the direction of the stretch is parallel to the applied field. This is true for both doped and undoped samples

(see Fig. 10). Why these magnetic moments are coupled to the CN network in such a way that they are predominantly oriented along the direction of the stretch is a noteworthy question in itself, but we will limit ourselves here only to the effects of this orientation.

A sample placed in magnetic field will then experience a stress. The resulting strain produces changes in resistance, mainly because of the changes in the energy gap. Theoretical and experimental studies have found that small axial strains can induce both band-gap openings and closings, which reflect in the electrical properties.[16-19] Simulations on semiconducting nanotubes of different chiralities connected in parallel[18] showed decreasing energy gap upon increasing strain, thus decreasing resistance with increasing strain, a trend consistent with experiment as well.[16] Similar simulations and experiments on metallic nanotubes found that their resistance increases with increasing strain.[18, 19] Moreover, the resistance of the joints between the nanotubes also may be affected by the strain.

If the $B$ field is aligned along the stretching axis (in our notation $B\parallel$ orientation for the laterally-stretched samples) the magnetic particles would exhibit the tendency to form chains. In ferrofluids where the magnetic particles are free to move the chain formation does take place. The particles attached to CN network do not really move, but the mutual interaction between the magnetic moments creates stress (predominantly compressive) which can release some of the original strain occurring from stretching. We expect thus that the overall axial strain along the stretching direction is reduced. If the positive MR is a result of localization, it should show the effect of increased energy gap upon decreasing strain. Indeed, the positive MR is enhanced when $B\parallel$ for the laterally-stretched sample. If $B$ is perpendicular to the plane of the fabrics, some strain might be generated along the thickness of the sample, but likely very small as the thickness is very small compared to the length or width of the sample. The effect is that $\gamma_{||} > \gamma_\perp$.

The effect of the stress on the negative term for the undoped samples is similar, i.e. $|\Delta R/R_{Dneg,lat,||}| > |\Delta R/R_{D,neg,lat,\perp}|$, but the MR is negative as the metallic CN's resistance decreases with decreasing the strain. In the case of the doped samples, the anisotropy is reversed, $|\Delta R/R_{Dneg,lat,||}| < |\Delta R/R_{D,neg,lat,\perp}|$, consistent more with weak localization, where the MR depends on to the area of the interfering loops perpendicular to the direction of the field. This is likely because the sample is degenerately doped and strain does not affect the resistance anymore.

## 5. CONCLUSIONS

Here we present results on the anisotropic properties of the magnetoresistance of CN sheets that have been partially-aligned by post-fabrication stretching, such that the direction of stretching was either parallel or perpendicular to the direction of applied electric current. The magnetic field has been applied either parallel or perpendicular to the plane of the sheet. As previously found for CN networks, the magnetoresistance is a superposition of two terms, one negative and one positive. Using high field fitting, we have separated the two terms and observed their anisotropy and dependence on the direction of stretching and doping. The main result is that the anisotropy of the two terms is different in magnitude, the anisotropy being more pronounced for the positive term. Moreover, we found that both terms are larger when a *B* field is applied parallel to the plane of the sheet compared to when it is applied perpendicular to the sheet for the undoped sample. Also, the MR is sensitive to the direction of alignment, as well as to the doping levels of the sheets. Analysis of our results within traditional theoretical frameworks like strong and weak localization indicates that the anisotropy is the result of a different effect. We propose a *B*-dependent strain cause, generated by the catalyst nanoparticles which are magnetized in the presence of magnetic field, producing strain which in its turn affects the value of resistance.

ACKNOWLEDGMENTS

The work at USNA was supported by the US Naval Academy Research Council. The authors would like to thank Nanocomp Technologies for providing the samples and M. White for useful discussions.

**Figure Captions:**

FIG. 1 SEM images of the doped longitudinally stretched sample. Inset: optical image of the sample with the copper plated pads. The arrow indicates the direction of stretching.

FIG. 2 The sheet resistance $R$ vs. the temperature $T$ for all four samples. Inset: the $T$-dependence of the resistive anisotropy with respect to the stretching direction. The cartoon is a drawing of the measurement configuration. White arrows indicate the stretching direction for the longitudinally-stretched samples (left) and laterally-stretched samples (right). The electrical current $I$ was always applied along the long axis of the samples. The magnetic field was applied parallel to the plane of the sheet or perpendicular to the plane of the sheet.

FIG. 3 (a) The sheet conductance $G$ vs. $T^{1/2}$ for the undoped samples. The straight line is a linear fit over limited temperature range. The inset shows the $T$-dependence of the reduced activation energy $W(T) = -\frac{d \ln \rho(T)}{d \ln T} = \frac{d \ln G}{d \ln T}$ for the laterally-stretched samples.

(b) The sheet conductance $G$ vs. $T^{1/2}$ for the doped samples. Inset: confirmation of the $ln(T)$ dependence up 60 K.

FIG. 4 The magnetic field $B$-dependence of the percentage magnetoresistance $\Delta R/R=[R(B)-R(0)]/R(0) \times 100$ for several representative temperatures for the laterally-stretched undoped sample (a) and doped sample (b).

FIG. 5 Example of the decomposing results of $\Delta R/R$ curves. The shown data correspond to the laterally-stretched samples, (a) undoped and (b) doped, at $T = 5K$. The curve labeled $\Delta R/R_{\parallel}$ represents the MR data in field applied parallel to the CN sheet. The curve labeled $\Delta R/R_{\perp}$ represents the MR data in field applied perpendicular to the CN sheet. The curves labeled $\Delta R/R_{\parallel,neg}$, and $\Delta R/R_{\perp,neg}$ represent the negative components under the two field orientations, while the curves labeled $\gamma_{\parallel}B^2$ and $\gamma_{\perp}B^2$ represent the positive components under the two field orientations. Inset: Comparison of the $B$-dependence of the negative terms for the two laterally-stretched samples, under both field orientations.

FIG. 6 The negative MR terms for the undoped laterally-stretched sample under the two field orientations and at $T = 10$ K vs. $B^{1/2}$. The solid curves are $B^2$ fits to the very-low $B$ regime. The magnetic field $B_\Phi$ where the fit deviates from the data was used to find the phase coherence length $L_\Phi$. The magnetic field $B_{sat}$ where the data curves saturate was used to find the size of the smallest interference loop, $l_{small}$.

FIG. 7 The $T$- dependence of the phase coherence length $L_\Phi$ for the (a) undoped and the (b) doped samples, under both field orientations. The straight lines are power fits to the data, $L_\Phi \sim aT^{-p/2}$. The resulting exponent for each curve is indicated in the legend.

FIG. 8 (a) The $T$- dependence of the saturation magnitude of the negative MR terms for the undoped samples, under both field orientations. (b) The temperature dependence of the $l_{small}$ for the undoped laterally stretched sample. Inset: Comparison of the $T$-dependence of the saturation quantities $B_{sat}$ and $|\Delta R/R_{neg,sat}|$.

FIG. 9 The *T*-dependence of the γ-coefficients, extracted from the MR positive terms $\Delta R/R_{pos} \sim \gamma B^2$, for the (a) undoped and (b) doped samples. Inset: The same data as in panel (a), but zoomed in the high temperature range.

FIG. 10 Pictures of the interaction between the undoped (a) longitudinally-stretched [(b) laterally-stretched] samples and the magnetic field between the poles of a horseshoe magnet. The samples lift to the top of the box, orienting such that the stretching axis is parallel to the direction of the field.

Figure 1.

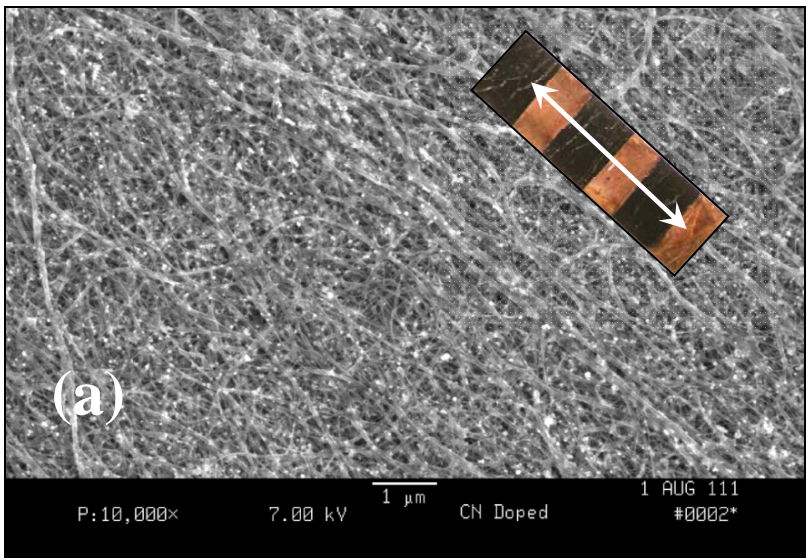

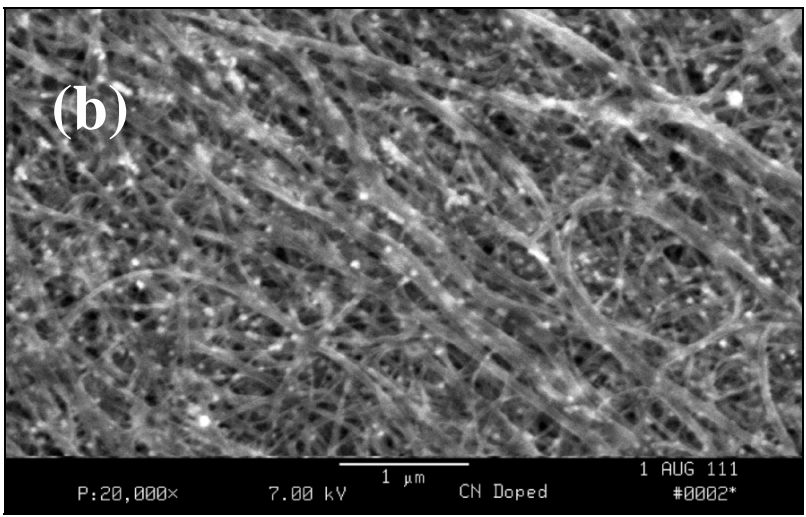

Figure 2.

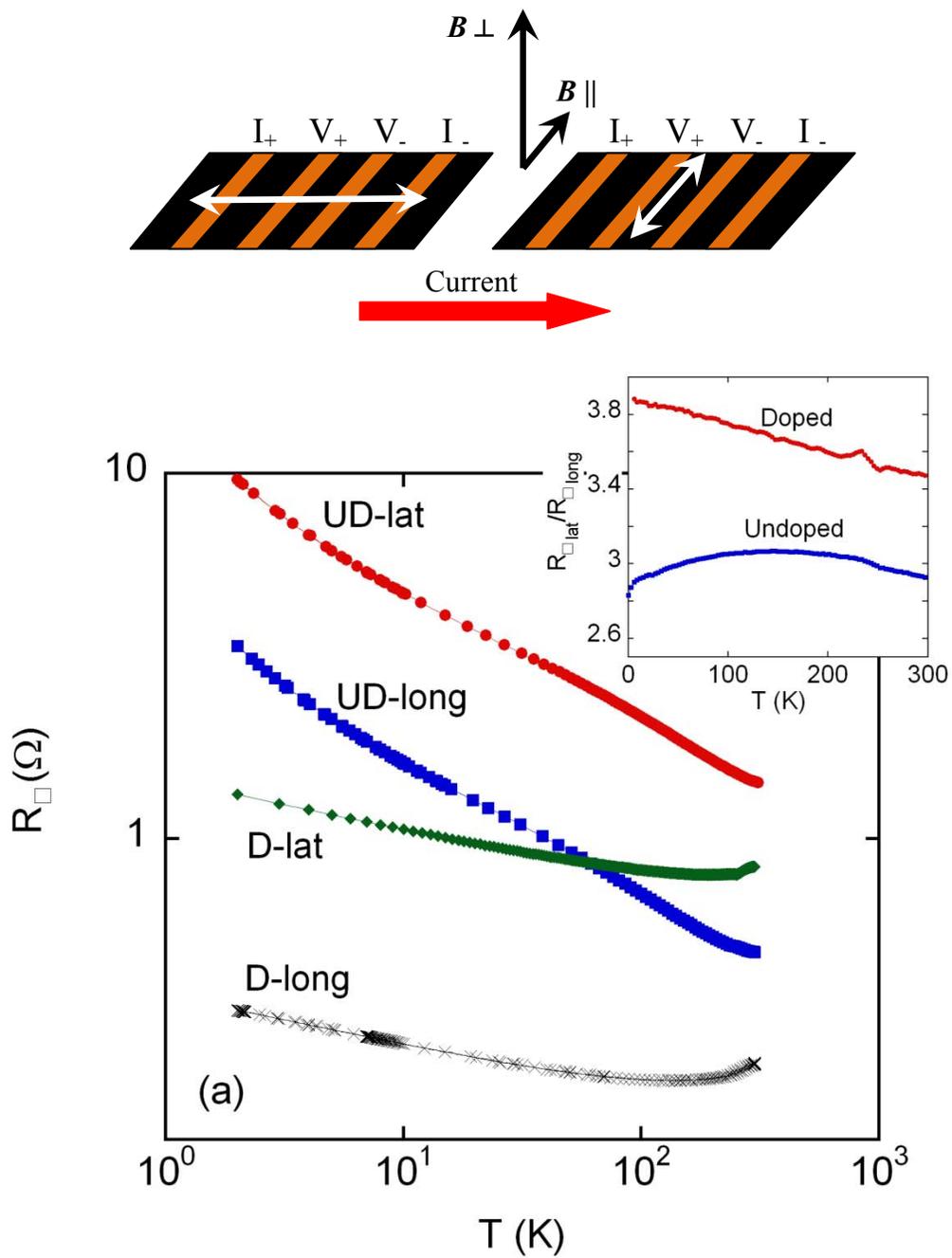

Figure 3

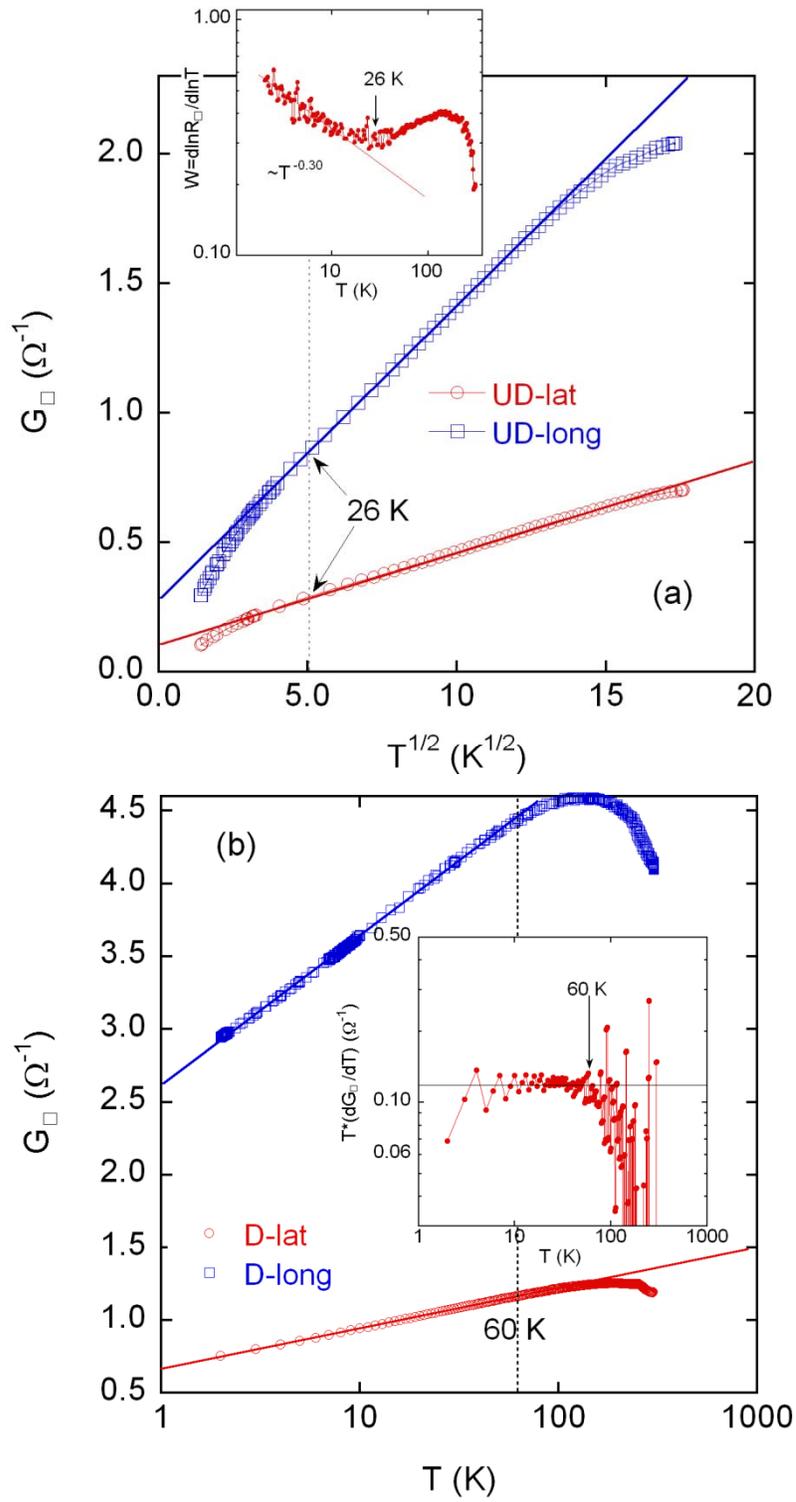

Figure 4

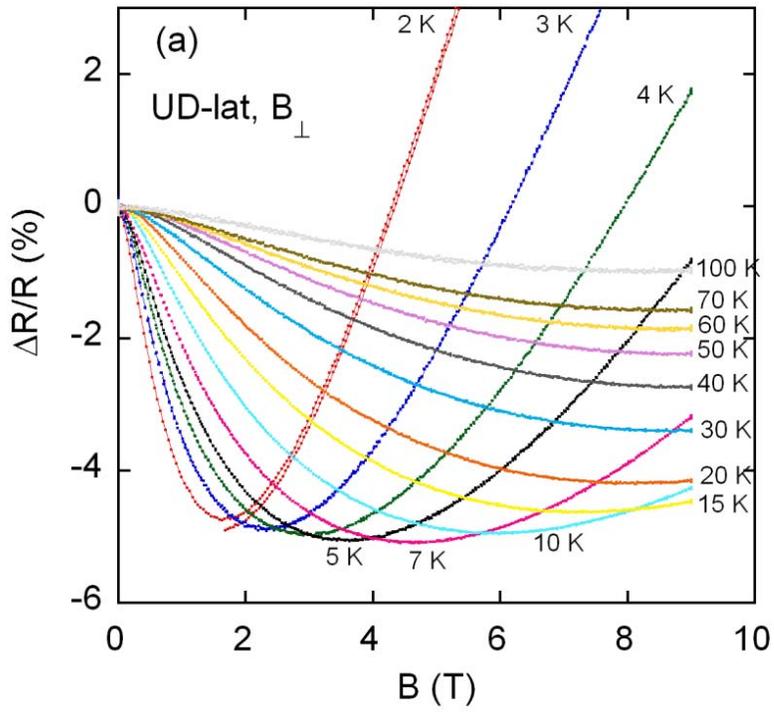

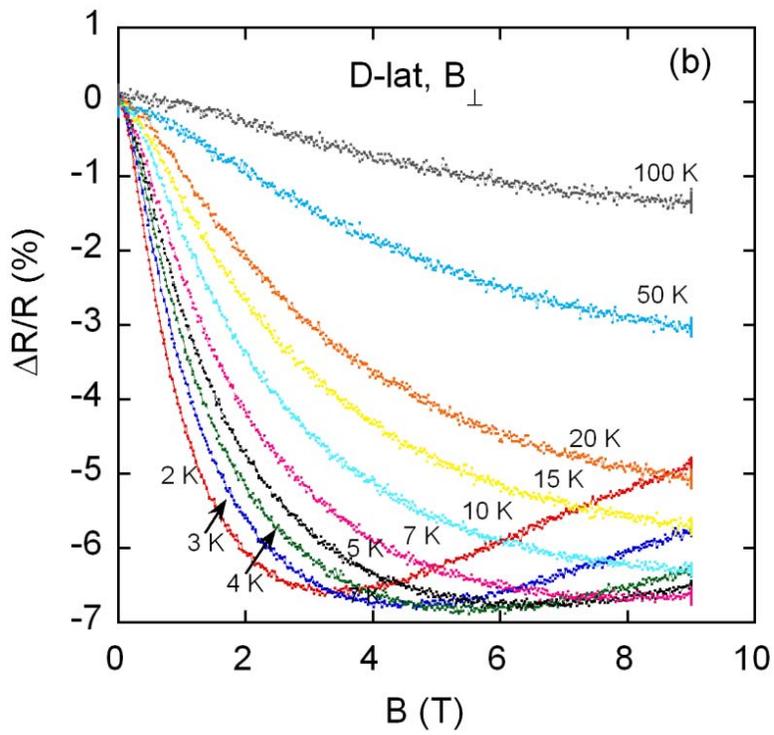

Figure 5

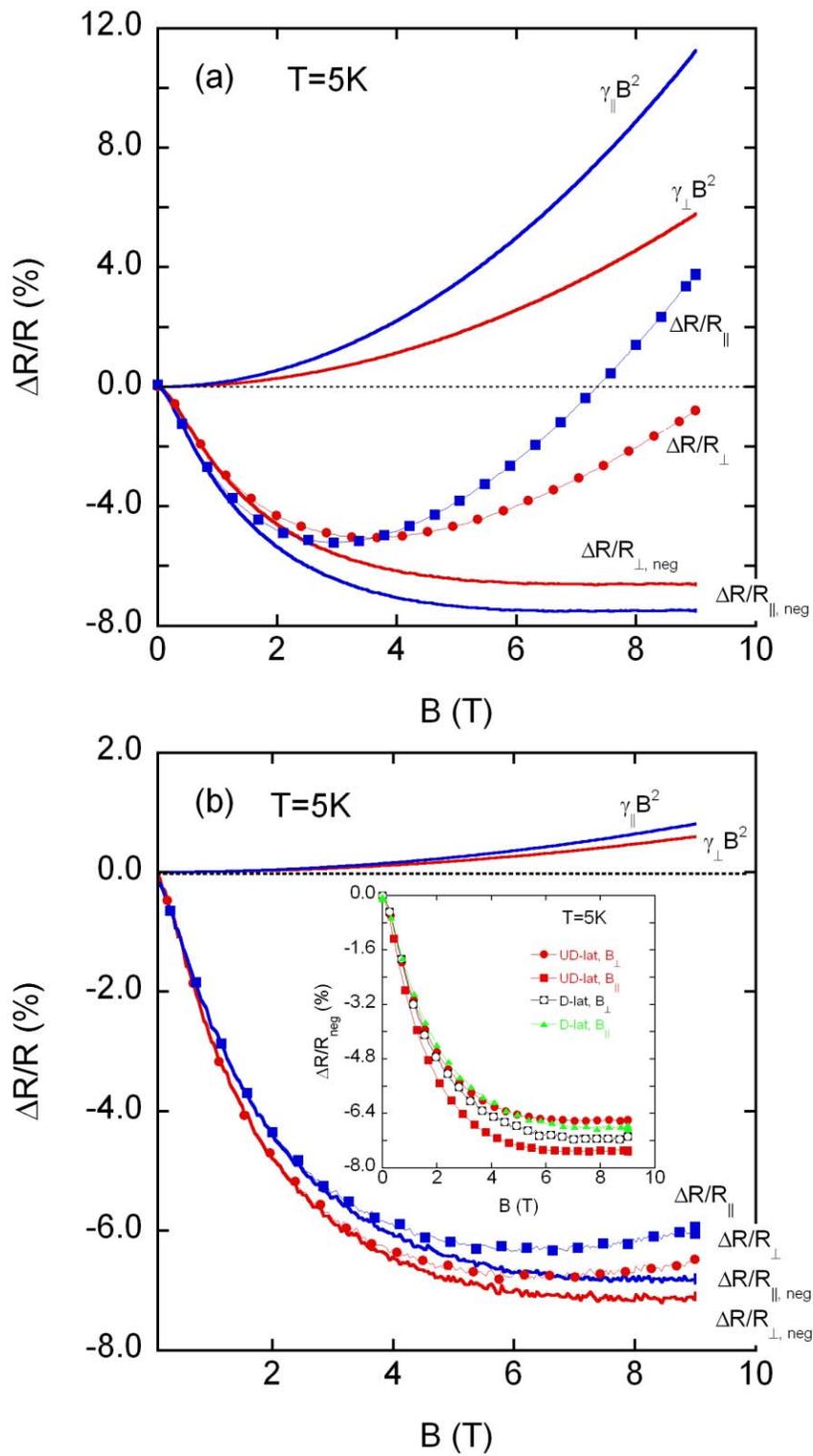

Figure 6

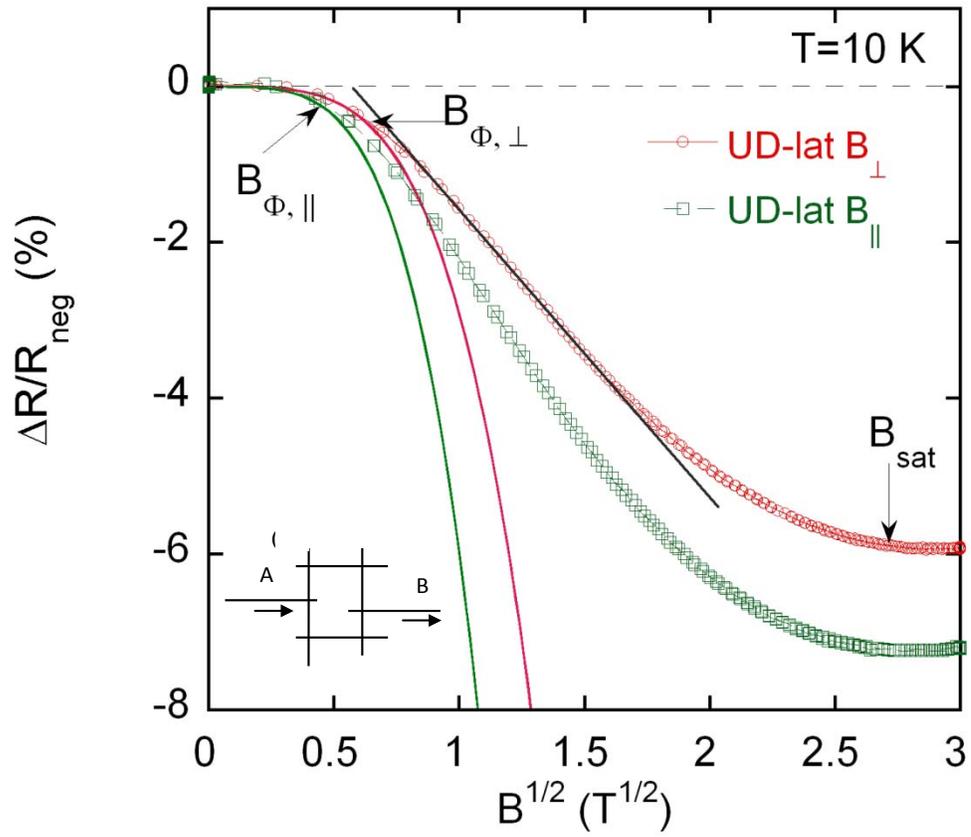

Figure 7

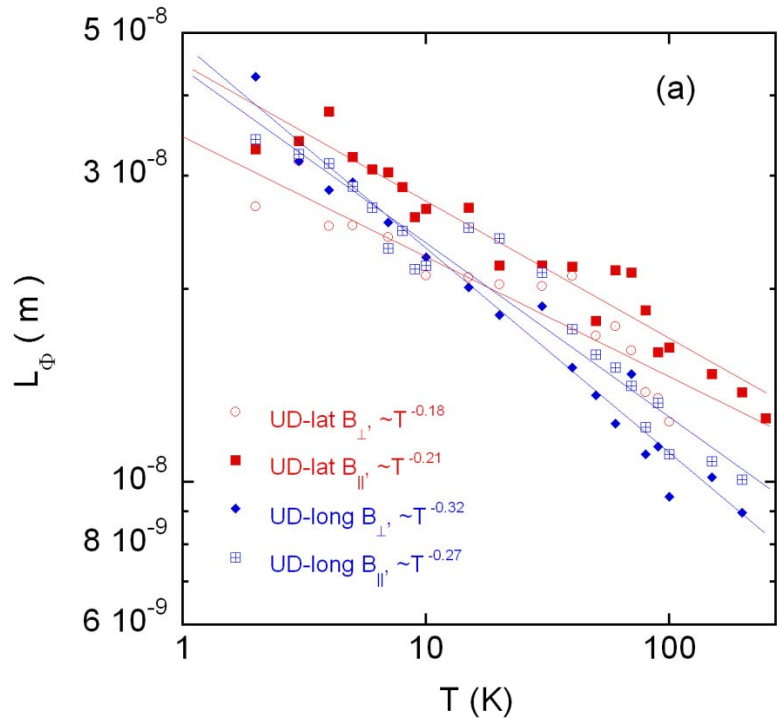

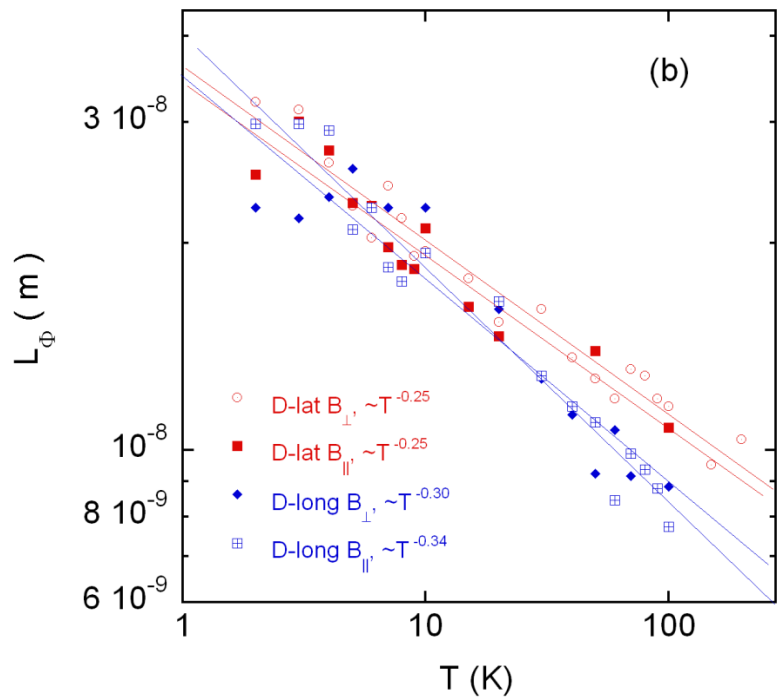

Figure 8

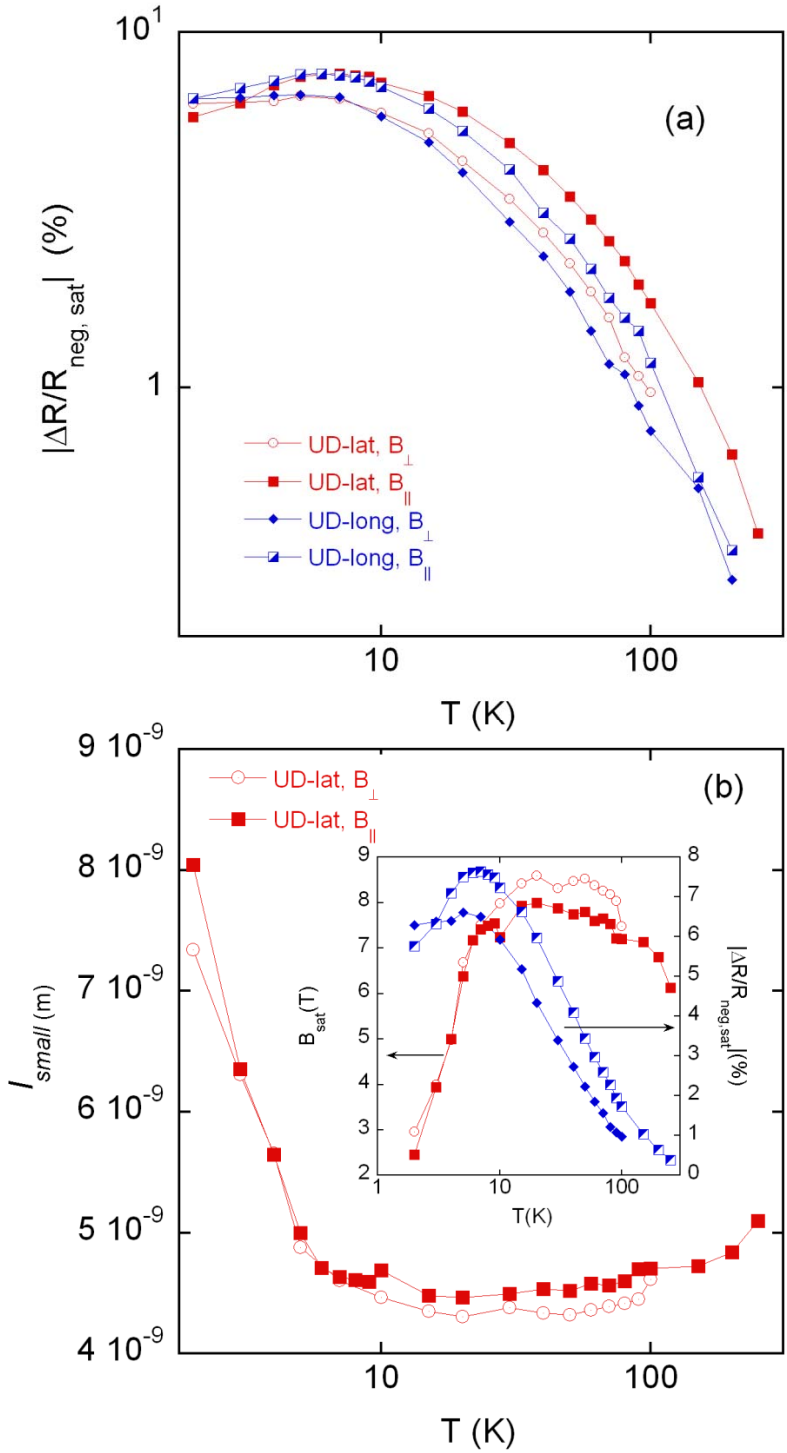

Figure 9

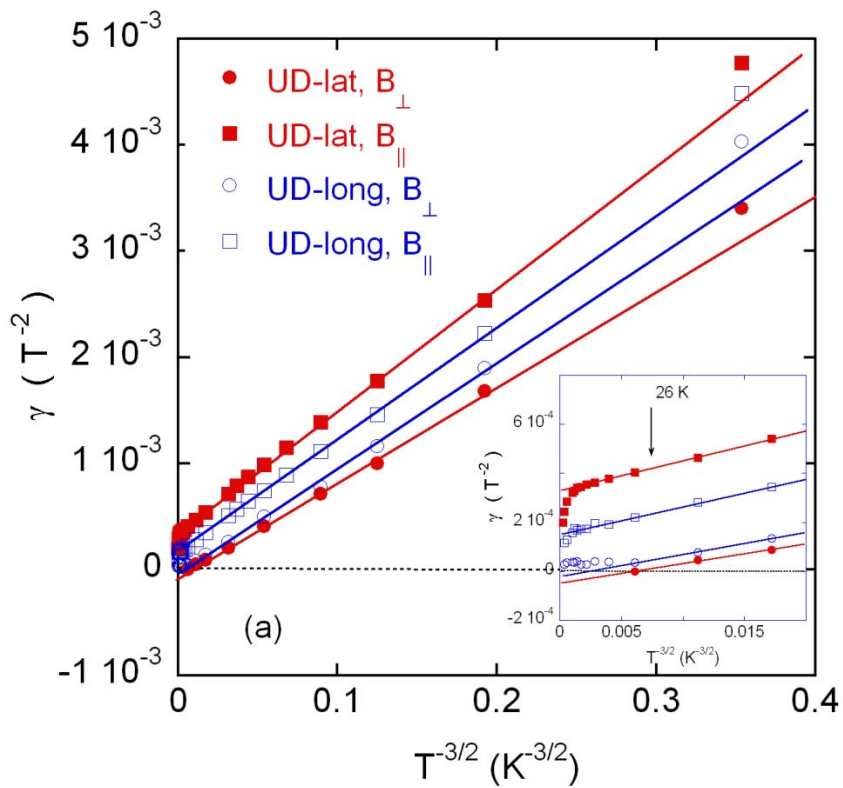

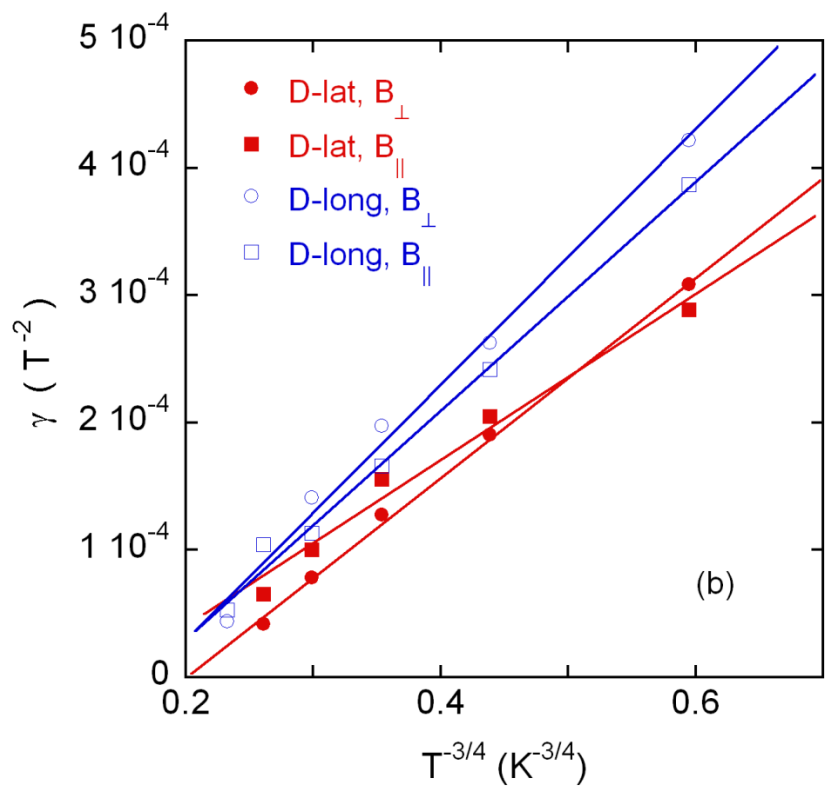

Figure 10

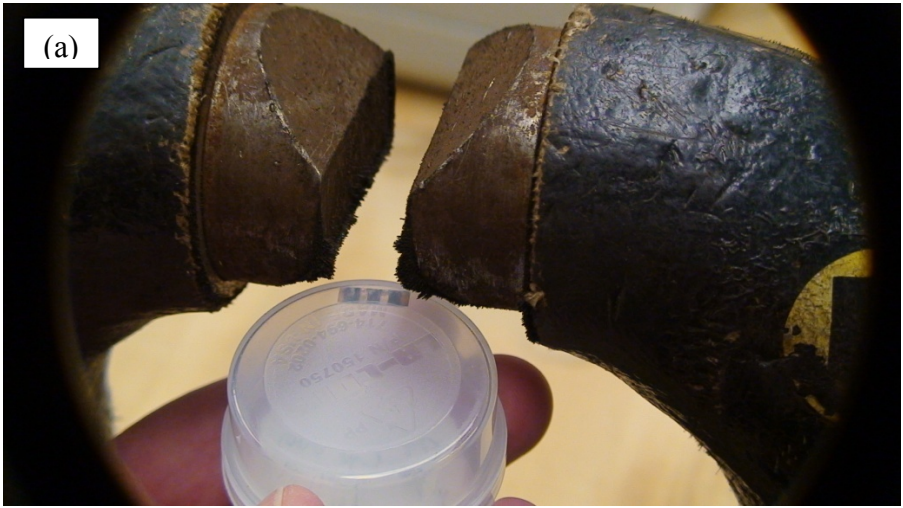

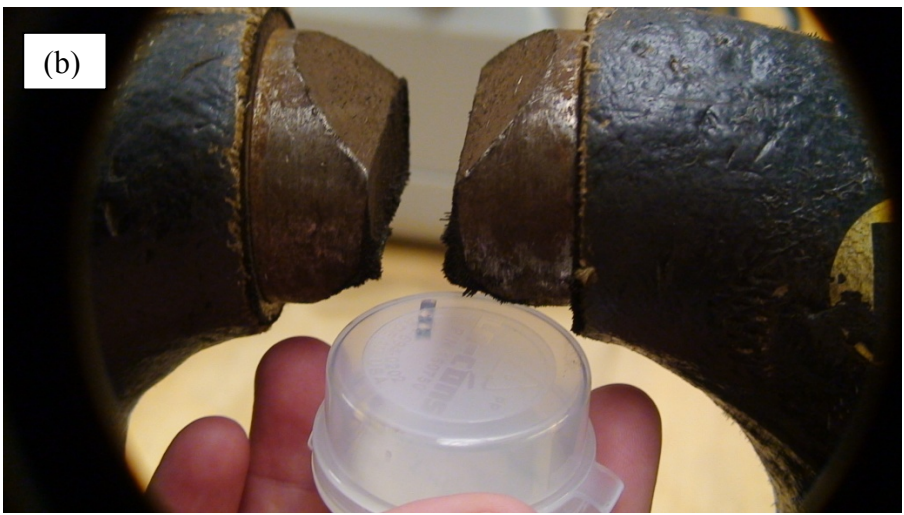